# Single Crystal Diamond Refractive Lens for Focusing of X-rays in Two Dimensions

S. Antipov[1]*, S.V. Baryshev[1], J.E. Butler[1, 2], O. Antipova[3], Z. Liu[4], S. Stoupin[4]**

[1] Euclid Techlabs LLC, Solon, OH 44139 USA

[2] Institute of Applied Physics of the Russian Academy of Sciences, Nizhny Novgorod, Russia

[3] Department of Biological and Chemical Sciences, Illinois Institute of Technology, Chicago, IL 60616 USA

[4] Advanced Photon Source, Argonne National Laboratory, Lemont, IL 60439 USA

Abstract:

We report the fabrication and performance evaluation of single crystal diamond refractive x-ray lenses with a paraboloid of rotation form factor for focusing x-rays in two dimensions simultaneously. The lenses were manufactured using a femtosecond laser micromachining process and tested using x-ray synchrotron radiation. Such lenses were stacked together to form a standard compound refractive lens (CRL). Due to the superior physical properties of the material, diamond CRLs are enabling and indispensable wavefront-preserving primary focusing optics for x-ray free-electron lasers and the next-generation synchrotron storage rings. They can be used for highly efficient refocusing of the extremely bright x-ray sources for secondary optical schemes with limited aperture such as nanofocusing Fresnel zone plates and multilayer Laue lenses.

* s.antipov@euclidtechlabs.com

** sstoupin@aps.anl.gov

The next generation light sources such as diffraction-limited storage rings and high repetition rate free electron lasers will generate x-ray beams with significantly increased peak and average brilliance. These future facilities will require x-ray optical components capable of handling large instantaneous and average power densities while tailoring the properties of the x-ray beams for a variety of scientific experiments.

Since the invention of x-ray compound refractive lenses (CRL) in the 90s [1] refractive x-ray optics has become one of the basic optical elements which couples the primary radiation of the highly brilliant x-ray sources with a variety of x-ray instruments to probe matter on the atomic level. Among the advantages of refractive x-ray optics are the preservation of the x-ray beam trajectory, robustness, compactness and relaxed requirements on surface quality owing to the transmission geometry. Commercially available polycrystalline beryllium refractive lenses with parabolic profiles [2, 3] are successfully used at the third-generation synchrotrons to refocus the x-ray source onto downstream optics, and, also as transfocators (reconfigurable stacks of CRLs) for selection of a photon bandwidth which is comparable to the bandwidth of high-brilliance undulator-based x-ray source [4, 5]. However, the choice of polycrystalline beryllium (dictated by very limited availability of high-quality single crystal material) results in limitations. These are the presence of grain boundaries which may distort the radiation wavefront, radiation damage limitations and environmental concerns (beryllium is highly toxic). The performance of a refractive lens can be improved by minimization of scattering from intrinsic inhomogeneities via a choice of material with low small angle scattering [2]. This also suggests preference of a single crystal over a polycrystalline material for lens fabrication. It is therefore imperative to develop the next-generation x-ray optics suitable for operation at the increased levels of x-ray power density.

Diamond has been the material of choice for several important high-heat-load applications in x-ray optics including double-crystal monochromators [6-8] and Bragg mirrors for the x-ray free-electron laser oscillator [9, 10]. Beyond that, it has been recently shown that diamond x-ray optics can withstand the direct x-ray free electron laser (XFEL) beam and provide the best performance parameters as Fresnel zone plates for nanofocusing of XFEL pulses [11], diffracting crystals for XFEL self-seeding [12, 13] and XFEL beam multiplexing [14, 15]. Theoretical studies (e.g., [16]) demonstrate that the present and future targets for instantaneous power densities are still below the radiation damage threshold in diamond. High average power density diamond resilience tests are currently in progress at the Advanced Photon Source [17].

Thus, the studies conducted so far demonstrate that diamond is an indispensable material for critical applications in x-ray optics for the present and the future light source facilities. A unique combination of exceptional intrinsic material properties such as high radiation hardness, record high thermal conductivity, small thermal expansion coefficient and uniformity of refractive index make single crystal diamond a material-of-choice for wavefront-preserving refracting optics in the critical *high-peak* and *high-average* power density applications. Several groups have reported successful fabrication and testing of diamond refractive lenses for focusing x-rays in one dimension (1D) [18-20]. These structures produced by ion/plasma etching have limited lateral apertures <=100 um, which is insufficient to accommodate the typical size of the high-brightness x-ray beams. Single crystal 1D diamond CRLs fabricated by laser micromachining which allow a suitable lateral aperture (~500 um) have been recently demonstrated [21].

In this work, we report the results of fabrication and characterization of single crystal diamond refractive lenses based on the established and widely accepted paraboloid of rotation form factor [2, 3]. It was shown that such lenses are free of aberrations and focus x-rays in two directions [3]. Femtosecond laser micromachining was used to produce the paraboloids of a small local radius of curvature (~105 μm) in single crystal CVD diamond plates available commercially. The fabricated lenses were tested using synchrotron radiation by reimaging a photon source of a bending magnet beamline (~198×78 μm$^2$) to a plane located at a distance of

54 m from the source. Nearly theoretical transmission over the aperture 230 μm in diameter was demonstrated at the photon energy of 13.6 keV. These lenses were then stacked together to form a standard CRL with ~ 3.36 m focal length and 450 μm aperture which was tested using undulator radiation at the photon energy of 11.85 keV. This CRL, placed at a distance of 62 m from an undulator, reimaged the 652 x 27 μm$^2$ source into a spot with a size of 52.6 x 21.4 μm$^2$.

While the choice of diamond based on its physical properties is obvious, the practical implementation is challenging. Conventional laser cutting by a standard nanosecond diamond cutter lasers provides non-satisfactory results caused by thermal fatigue in diamond [22]. Unlike the conventional approach, femtosecond laser pulse duration is extremely short: the material is ablated and pulsed heating effects are minimized. The lenses presented here were manufactured from single crystal CVD diamonds of various grades with thicknesses of about 500 μm. To machine a 2D paraboloid, the fs-laser beam was steered by a galvo mirror to ablate circle patterns gradually reducing the circle diameter with depth. The largest diameter on the diamond surface was about 450 μm. Identically, a matching paraboloid was micromachined on the opposite side of the diamond plate (Figure 1). After micromachining diamonds were cleaned in a hot mineral acid bath containing an oxidizing agent.

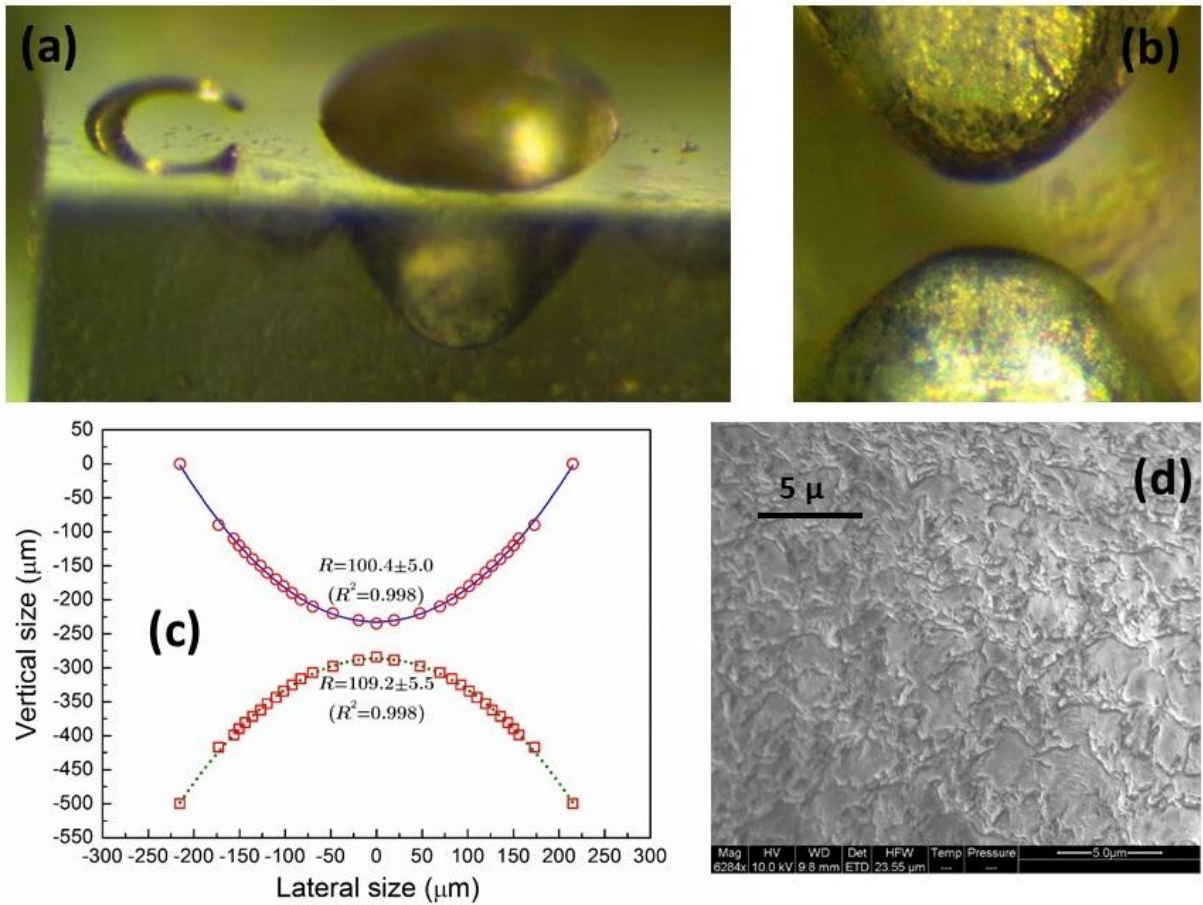

Figure 1. Microscope image of the lens at various angles (a, b). (c) Profile of the lens measured by an optical profilometer and parabolic fit with radii of curvature. (d) Scanning electron micrograph of the lens surface shows roughness on the order of 1 μm.

It was experimentally established, that about 40 μm distance (waist) can be left between the paraboloids. The diamond would get punctured when a smaller paraboloid separation was attempted. Figure 1 (a, b) shows the lens profile image taken by an optical microscope. The parabolic shape can be seen through the transparent side of the diamond plate (Fig. 1a). Two parabolas are depicted on Figure 1 (b), an image taken from the side of the diamond plate. When the first paraboloid was machined on the diamond surface, the total depth of the cut was measured. The second paraboloid was machined with the laser power scaled to maintain the separation of at least 40 μm. This is why the paraboloids had different radii of curvature (100 and 109 μm) as measured by white light interferometry (WLI) using an optical profilometer MicroXAM-1200 [23]. WLI could not automatically reconstruct 3D profile of the lenses due to poor reflectivity in the direction normal to the surface and high diamond transparency. By combining objectives of different magnification and enhanced contrast imaging, interference fringe diameters at different vertical positions of the focal plane, with the diamond surface set to zero as a reference, were recorded. All fringes (representing 2D cross-sections of a 3D solid at every focal plane parallel to the surface) were nearly circular confirming that the micro-machined lenses had rotation symmetry. Finally, curvature radii were measured via plotting a dependence of the fringe diameter on depth (symbols in Fig.1c) and fitting this dependence with a quadratic function (lines in Fig.1c). Goodness of determination ($R^2$) was 0.998.

The measured waist between the paraboloids in the tested lenses was 50 μm. Scanning electron microscopy image of the lens surface (Fig.1d) showed roughness of about 1 μm r.m.s.

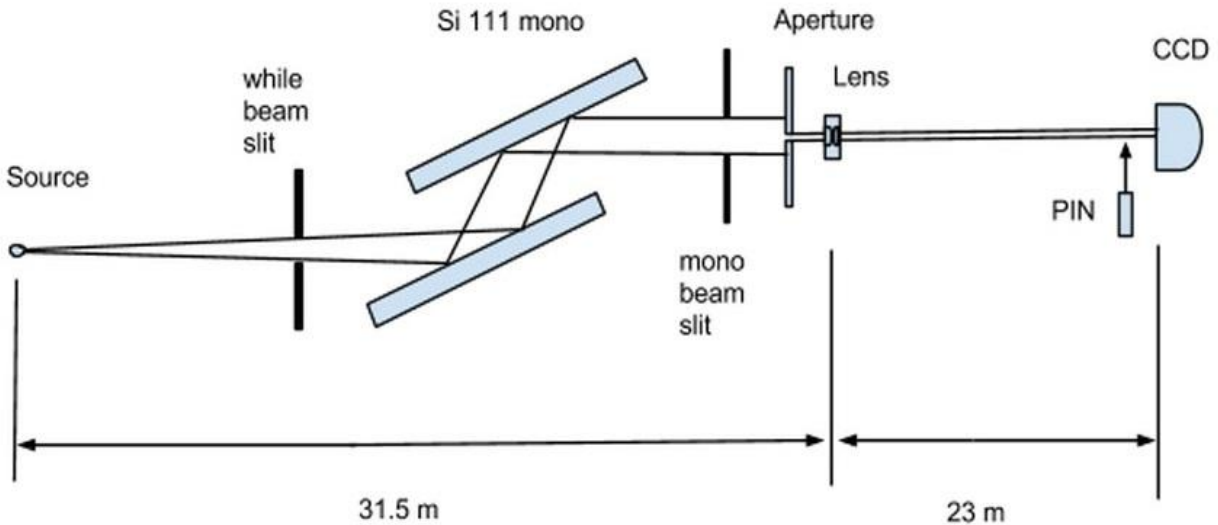

Figure 2. Experimental setup (see text for details).

The first experiment was performed at 1-BM Optics testing beamline of the Advanced Photon Source. The experimental arrangement is shown in Fig.2. A Si 111 double crystal monochromator was tuned to a photon energy of 13.6 keV. To limit the wavefront distortion caused by the heat load on the monochromator first crystal, the size of the incident white beam was set to 1×1 mm$^2$ using the upstream white beam slit. The size of the monochromator exit beam was selected to be comparable with the total lens aperture using mono beam slits. The lenses were placed at a distance of 31.5 m from the source (bending magnet). A circular pinhole aperture with diameter ~230 μm was placed in front of the lens. Imaging of the transmitted/refracted beam profile was performed using an area detector (CCD) placed at ~23 m downstream of the lens. A retractable solid state pin diode detector (PIN) was used for relative measurements of the lens transmission (lens in/out of the beam). The area detector was equipped with 250-μm-thick LUAG:CE scintillator and an objective lens with 10X magnification. The field of view for the area detector was 1.6×1.4 mm$^2$ with the effective pixel size of ~0.65 μm.

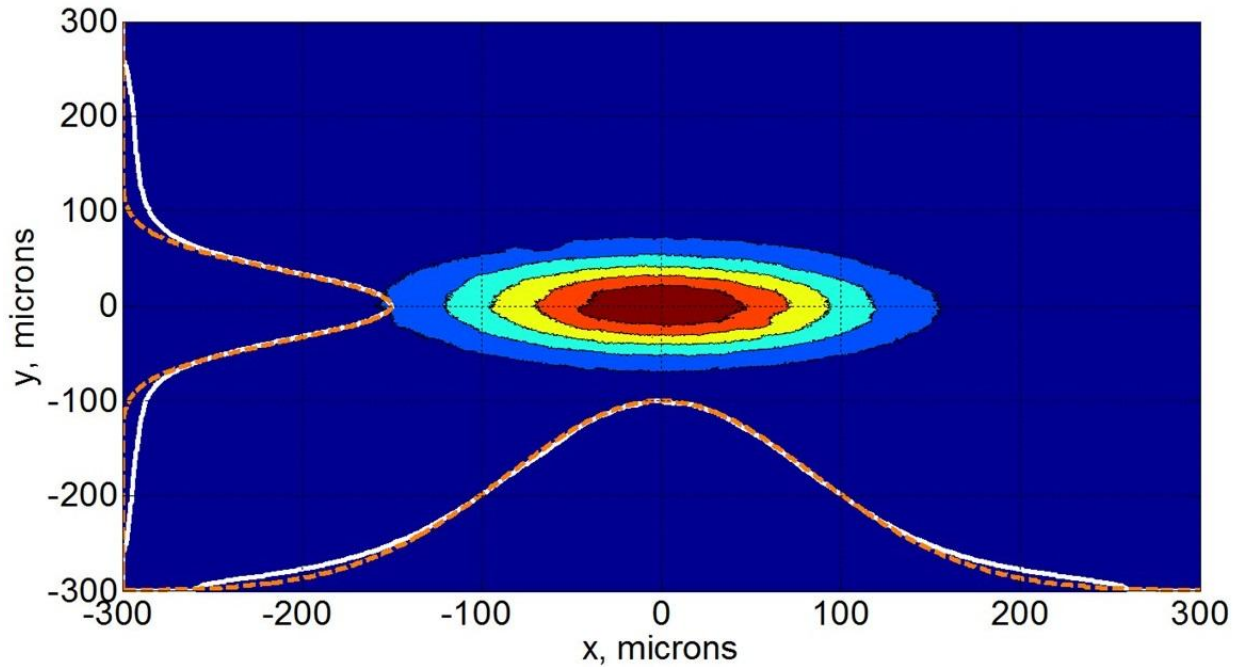

Figure 3. Image of the 13.6 keV x-ray beam focused by a single lens. Beam projections on *x* and *y* axes (white, solid), Gaussian fit (orange, dashed). The refocused source size is 183 μm x 86 μm (FWHM).

In the experiment the bending magnet source size was reimaged to the far experimental station at the beamline while the lens was placed in the near station. The nominal demagnification ratio was 0.73. The bending magnet source size was 198(4)×78(1) μm$^2$ full width at half maximum (FWHM). These values were taken as equal to the corresponding electron beam sizes of the bending magnet source which were extracted from the log of synchrotron parameters.

Few different diamond lenses were characterized. These samples were made from several different single crystal CVD diamond grades. No substantial difference in the lens performance parameters was found for any of the lenses in the two samples. To arrive to the optimal refocusing geometry the photon energy of the monochromator was tuned to obtain a minimal vertical size of the source in the imaging plane located at the fixed distance corresponding to the focal distance f = 13.3 m. The optimal photon energy was found to be 13.6 keV. The equation for the focal distance of the lens f = R/(2*δ(E)) allows us to evaluate the effective radius of curvature R since the refractive decrement δ(E) is a tabulated value. The effective radius of curvature was found to be R = 105 µm, which is in good agreement with the results by optical profilometry (Fig. 1c).

In this optimal condition we measured the source image size by fitting a 2D Gaussian distribution to the intensity distribution in the imaging plane and transmission of the lens using the integrating detector. In order to compare experimental results with the theory the measured transmission value T = 0.973 was renormalized by the fraction of the signal forming the Gaussian portion of the image. The obtained effective transmission $T_{eff}$ was 0.87. The transverse distribution of the beam obtained with the area detector and the Gaussian fit projections are shown in Fig. 3. The measured and theoretical performance parameters are summarized in Table 1.

A relevant theoretical model describing transmission through the lens of material with refractive index $n = 1 - \delta - i \cdot \lambda\mu/(4\pi)$, where $\lambda$ is the radiation wavelenth, takes into account the interface roughness, σ. The effective transmission is given by [3]:

$$T = \frac{\exp(-\mu d)}{2a_p}[1 - \exp(-2a_p)] \quad (1)$$

The attenuation factor $a_p$ is given by:

$$a_p = \frac{R_0^2}{2R^2}[\mu R + 2Q_0^2\sigma^2], \quad (2)$$

where $Q_0 = 2\pi d/\lambda$, is the momentum transfer for transmission through an air-lens interface at normal incidence. The distance between the lenses paraboloids is $d$.

This equation allows us to evaluate the transmission of the lens taking into account the surface roughness, *σ = 1.0 µm,* estimated from the SEM measurements. The obtained value is *T = 0.89* which agrees with the effective measured transmission $T_{eff}$ = *0.87*.

The gain of a refractive lens is defined as the ratio of the intensity in the focal spot to the intensity behind a pinhole (*2R₀ = 230 um*) equal to the spot of the lens: $g = T_{eff} \cdot 4 \cdot R_0^2 / (B_v \cdot B_h)$ [3]. An ideal lens in the geometry shown in Fig.2 should refocus the source to an image with sizes $B_h$ = *145 µm* and $B_v$ = *57 µm*, both FWHM. The obtained values are larger (183 µm x 86 µm) which can be attributed to deviations of the lens profile from the ideal shape.

**Table 1.** Single lens performance parameters.

| E [keV] | T, theory | Gain, theory | $T_{eff}$, experiment | $B_h$, measured | $B_v$, measured | Gain, exp |
|---|---|---|---|---|---|---|
| 13.6 | 0.89 | 5.71 | 0.87 | 183 µm | 86 µm | 2.83 |

We observe a difference in the measured and the theoretical gain similar to what was reported earlier in a one dimensional diamond lens [21]. The laser micromachining process featuring a step by step ablation of material produces a modulated parabolic profile that results in broadened focal spot and reduced gain [21].

After single lens measurements we assembled a CRL prototype by stacking together three diamond lenses. The total device size length was 1.5 mm with the focal length of 3.36 meters at 11.85 keV x-ray energy and a total aperture of 450 μm. The CRL was tested using undulator radiation at the Biophysics Collaborative Access Team at the APS [24]. It was installed at 62 meters from the source. A circular pinhole aperture with a diameter ~ 400 μm was placed in front of the CRL. An ideal CRL would refocus the nearly Gaussian undulator source with a size of 652 x 27 μm$^2$ to an image with a size 37.3 x 1.55 μm$^2$ FWHM [3]. The focused beam profile measured by CCD at 3.55 m from the lens was 52.6 x 21.4 μm$^2$. This discrepancy is attributed to the precision of the lens stacking. The horizontal size of the beam is significantly larger than the vertical size and therefore is much less sensitive to CRL stacking misalignments.

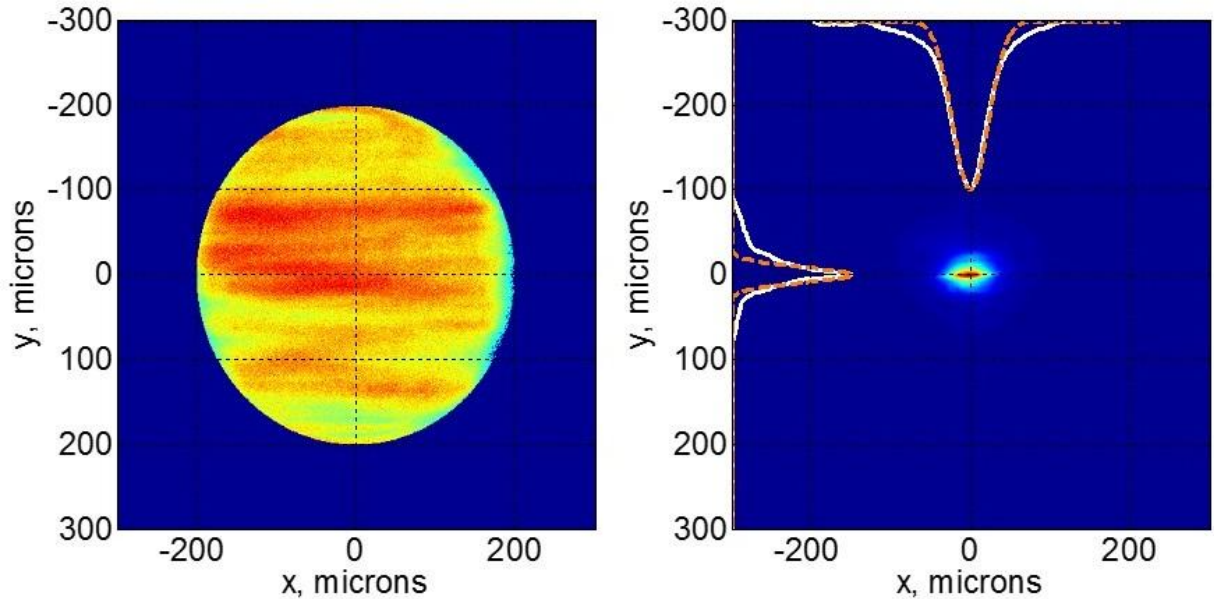

Figure 4. Left (lens out): Image of the aperture limited beam. Right (lens in): Image of the 11.85 keV x-ray beam focused by the CRL. Beam projections on *x* and *y* axes (white, solid), Gaussian fit (orange, dashed). The focused beam size is 52.6 x 21.4 μm$^2$.

The transmission of the CRL was 74%. The focused beam deviates from the Gaussian due to imperfections of lens geometry, limited stacking precision and the surface roughness. However the Gaussian portion of the beam is 71%. To calculate the gain of the lens we corrected the experimental transmission value to include only the Gaussian portion. The effective transmission of the CRL was 0.53 and the corresponding gain was 53.5. The prototype of a 2D diamond crystal CRL is demonstrated to be functional for refocusing of an undulator source onto secondary optical schemes with reasonably small non-Gaussian contributions. The analysis of the measured parameters indicates a straightforward approach to achieve improvement in performance (e.g., precise lens stacking and further optimization of micromachining procedures).

In summary, a two-dimensional x-ray refractive lens was fabricated out of single crystal diamond by fs-laser micromachining and experimentally tested. Efficient refocusing of a bending magnet source to a nearly Gaussian beam profile was demonstrated at a photon energy of 13.6 keV. Three of such lenses with average radii of curvature of 105 μm were stacked together to form a compound refractive lens. This compound lens was tested with an undulator beam at 11.85 keV photon energy. The undulator source was focused into a spot with a size of 52.6 x 21.4 μm$^2$. The measured gain of the CRL was 53.5 which renders it suitable for moderate focusing applications of high-power-density x-ray beams. Further improvements of the CRL performance are anticipated to be achieved by implementation of precise stacking schemes and refinement of laser micromachining and surface polishing procedures. Diamond CRLs will become the main choice for x-ray focusing of primary beams at future light sources. Due to outstanding thermal and radiation hardness properties of diamond these devices can withstand extremely high instantaneous flux densities generated by XFELs as well as unprecedented average flux densities of the next-generation light sources with high repetition rates.

We are indebted to K.-J. Kim, Yu.V. Shvyd'ko, C. Jacobsen and A. Sandy for helpful discussions on the topic of x-ray refractive optics. R. Woods and K. Lang are acknowledged for technical support. Euclid Techlabs LLC acknowledges support from DOE SBIR program grant No. DE-SC0013129. Use of the Advanced Photon Source was supported` by the U. S. Department of Energy, Office of Science, Office of Basic Energy Sciences, under Contract No. DE-AC02-06CH11357. J.B. acknowledges the support of the Act 220 of the Russian Government (Agreement no. 14.B25.31.0021 with the host organization IAP RAS). BioCAT acknowledges support by grant 9 P41 GM103622 from the National Institute of General Medical Sciences of the National Institutes of Health."